\documentstyle[seceq,mbf,wrapft,epsf]{ptptex}

\pubinfo{}  
\preprintnumber[3cm]{
KUNS-1325\\PTPTeX ver.0.8\\ August, 1997}

\title{T-odd quark fragmentation function and 
transverse spin\\ asymmetries in the pion production
}

\author{Katsuhiko {\sc Suzuki}
}

\inst{RCNP, Osaka University, Osaka 567-0047, Japan
}

\abst{
We study the time-reversal odd quark fragmentation function and its  
consequences on the hard processes. 
T-odd quark fragmentation function may arise, although QCD is 
T-invariant, from the 
non-perturbative dynamics in the fragmentation process.  
Assuming the factorization, we extract the T-odd fragmentation function 
from the inclusive pion production in the transversely polarized 
proton-proton collision.  
We then estimate the single spin asymmetry $A_{OT}$ of the semi-inclusive 
deep inelastic scattering with unpolarized lepton and the transversely 
polarized proton.  
We also discuss the meson cloud effects on the transversity 
distribution $h_1(x)$ of the nucleon.  
Asymmetry for the $\pi^-$ production is found to be much smaller than 
the naive expectation, since the 
$d$-quark transversity is considerably suppressed by the pion cloud effects.  
}

\begin{document}

\maketitle


\makeatletter
\if 0\@prtstyle
\def\asp{.3em} \def\bsp{.26em}
\else
\def\asp{.3em} \def\bsp{.3em}
\fi \makeatother

\section{Introduction}

Recently much attention is paid on the transverse spin phenomena at high 
energies.  
Large single spin asymmetries of the pion production cross section are 
found 
in the transversely polarized proton-proton collision 
$\vec p + p \to \pi^a + X$\cite{pp}.  
Similar results are obtained for the kaon and eta meson productions.  
Significant polarizations of hyperons are also measured in the unpolarized 
hadron-hadron collision.  
Since the hard scattering process calculated by perturbative QCD predicts 
negligible asymmetry $\sim \alpha \frac{m}{Q}$\cite{Repko}, we expect some 
non-perturbative mechanism 
as the origin of such transverse spin effects.

Collins pointed out that the non-trivial azimuthal angle dependence of the 
distribution of  
produced particles exists in the fragmentation of the transversely polarized 
quark\cite{Collins,Mulders,Boer}.  
Let us consider the pion production as an example. 
The `standard' unpolarized quark fragmentation function for the pion is 
defined by
\begin{eqnarray}
D_{\pi / a}(z,k_\bot )&=&\sum\limits_X {\int {{{dy^-d^2y_\bot } 
\over {12(2\pi )^3}}}}\,e^{ik^+y^--ik_\bot \cdot y_\bot } \nonumber \\
&&\times \mbox{tr} \; \gamma ^+\left\langle {0|\psi _a(0,y^-,y_\bot )|h,X}
 \right\rangle \left\langle {h,X|\bar \psi _a(0)|0} \right\rangle
\label{def1}
\end{eqnarray}
where a pion with momentum $p$ and $z = p^+ / k^+$ 
is created by the quark $a$ with momentum 
$k$.  
When the quark is transversely polarized, it is possible to consider the 
following one, 
\begin{eqnarray}
H_{1 \; \, \pi / a}^\bot (z,k_\bot , s_\bot)&=&\sum\limits_X 
{\int {{{dy^-d^2y_\bot } 
\over {12(2\pi )^3}}}}\,e^{ik^+y^--ik_\bot \cdot y_\bot } \nonumber \\
&&\hspace{-1.3cm}
\times \mbox{tr} \; \gamma ^+\gamma _5\kern 1pt \gamma _\bot \cdot s_\bot 
\left\langle {0|\psi _a(0,y^-,y_\bot )|h,X} \right\rangle 
\left\langle {h,X|\bar \psi _a(0)|0} \right\rangle
\label{def2}
\end{eqnarray}
where $s_\bot$ is the transverse spin of the quark.  
This fragmentation function is proportional to 
\begin{eqnarray}
\epsilon_{\kappa \lambda \mu \nu}s_\bot^\kappa k_\bot^\lambda 
p^\mu n^\nu
\end{eqnarray}
where $n$ is the light-cone vector.  
Time-reversal invariance prohibits existence of such a quantity, 
unless there are non-trivial phase differences of the amplitudes  
from the final state interaction in the hadronization process.  However, 
non-perturbative hadronization process may generate such phase 
differences and allow a T-odd quark fragmentation\cite{Collins,JJT}.  
We will demonstrate how the T-odd fragmentation process is generated using 
a toy model in section 2.

If this is the case, one can use the T-odd quark fragmentation process in 
order to probe the chiral-odd transversity distribution function $h_1(x)$ 
of the nucleon\cite{JJ}.  
The transversity distribution $h_1(x)$ has not been measured so far, because 
it can not be probed by the deep inelastic scattering due to its chiral-odd 
nature.   
If we assume the existence of the T-odd fragmentation function and the 
factorization, we may write the single spin asymmetry of the hard process 
e.g.~semi-inclusive 
deep inelastic scattering with the transversely polarized nucleon 
and unpolarized lepton as
\begin{eqnarray}
\sim h_1(x) \otimes \sigma _h \otimes H^\bot \; .
\label{factor}
\end{eqnarray}
This unable us to measure the transversity distribution $h_1(x)$ 
by the hard processes. 
Here, we focus on the single pion production in the 
proton-proton collision and the semi-inclusive deep inelastic scattering 
off the proton.    
In both cases a target proton is transversely polarized.  
In section 3 we try to extract the T-odd fragmentation function from the 
existing data 
of the transversely polarized proton-proton collision in which 
significant analyzing powers are observed\cite{pp}.  
After fixing the fragmentation functions, we analyze the semi-inclusive 
DIS with a unpolarized lepton and a transversely polarized proton 
$\ell + \vec p \to \ell' + \pi^a + X$ in section 4.  
We shall estimate the transverse spin asymmetry which will be accessed by 
future experiments.

In section 5 we emphasize roles of the meson clouds in the transversity 
distribution function.  
Physical nucleon state involves meson clouds, $\pi, K \dots$ as higher 
Fock components.  Existence of such meson clouds is strongly suggested
by the large isospin symmetry breaking in the nucleon sea.  
In ref.~\cite{KW} we have found that the pion cloud gives quite different 
contributions to the transversity distribution $h_1(x)$ and the 
helicity one $g_1(x)$.  
Such interesting differences will be checked by forthcoming experiments, 
and provides new insight of the roles of the meson clouds.

\section{Model of T-odd fragmentation function}

Let us consider how T-odd fragmentation function appears by using a toy 
model.  
For this purpose, we take Nambu$-$Jona-Lasinio model\cite{NJL} in which 
non-perturbative gluon degrees of freedom are frozen into the effective 
4-quark interaction.  This 
model provides a schematic picture of the dynamical chiral symmetry 
breaking in QCD, and describes the Goldstone pion as a highly collective 
$q \bar q$ state.

\begin{figure}
\begin{minipage}[c]{60mm}
            \epsfxsize = 4cm   
            \centerline{\epsfbox{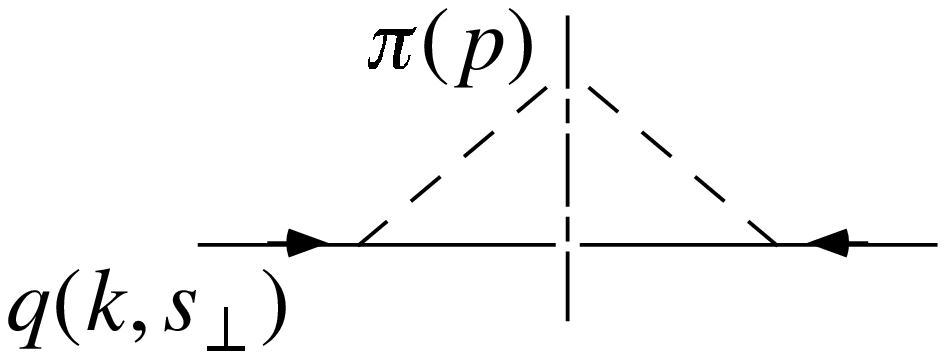}}
          \caption{Tree diagram for $q \to \pi$ process.  Solid and 
dashed lines represent the quark and pion, respectively.  }
\end{minipage}
\hspace{\fill}
\begin{minipage}[c]{70mm}
            \epsfxsize = 6cm   
            \centerline{\epsfbox{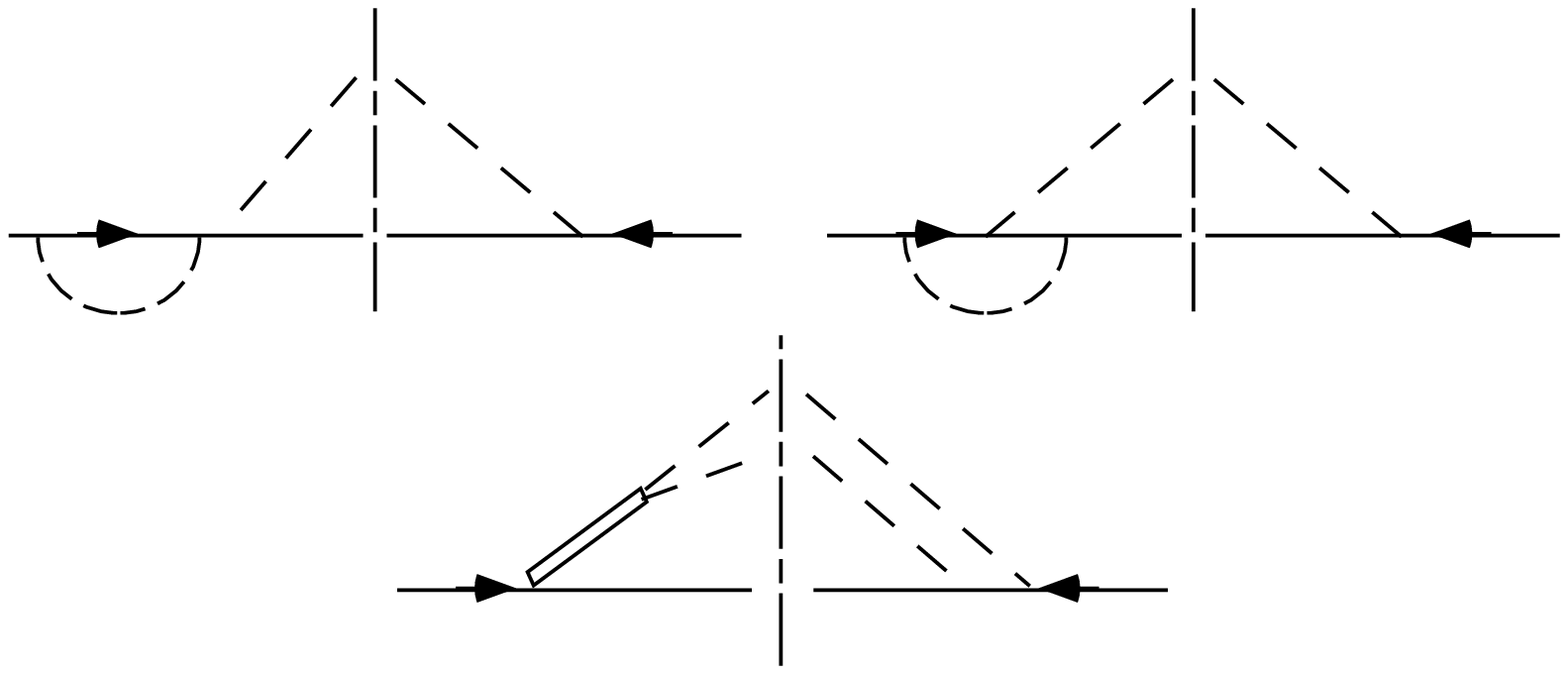}}
          \caption{Typical diagrams contributing to T-odd process.  
Thick line denotes the $\sigma$ meson}
\end{minipage}
\end{figure}

Evaluating the matrix element in eq.~(\ref{def1}), one finds the unpolarized 
quark fragmentation function for the pion at tree level (Fig.1) as
\begin{eqnarray}
D(z,k_\bot)&=&g^2\int {{{dk^-} \over {16\pi ^4}}}{1 \over {\left( {k^2-m^2} 
\right)^2}}2\pi \delta \left[ {(k-p)^2-m^2} \right] \nonumber \\
&&\times {1 \over 2} \mbox{tr} \left[ {\gamma ^+(\not k+m)\gamma ^5
(\not k-\not p+m)\gamma ^5(\not k+m)} \right] \nonumber \\
&=&{{g^2} \over {8\pi ^3}}{z \over {1-z}}\left[ 
{{1 \over {\left( {m^2-k^2} \right)}}+{1 \over z}{1 \over {\left(
 {k^2-m^2} \right)^2}}} \right]
\end{eqnarray}
where $m$ is the constituent quark mass and 
$k^2 = z/(1-z) k_\bot^2 + m^2 /(1-z) + m_\pi^2 / z$.

On the other hand, it is obscure whether or not T-odd fragmentation function 
exists in the hadronization process.  
Tree diagram calculation of eq.~(\ref{def2}) yields 
\begin{eqnarray}
H_1^\bot (z,k_\bot,s_\bot)&=&g^2\int {{{dk^-} \over {16\pi ^4}}}
{1 \over {\left( {k^2-m^2} \right)^2}}2\pi \delta \left[ {(k-p)^2-m^2} 
\right] \nonumber \\
&&\times {1 \over 2}\mbox{tr} 
\left[ {\gamma ^+\gamma ^5s_\bot \cdot \gamma _\bot 
\,(\not k+m)\gamma ^5(\not k-\not p+m)\gamma ^5(\not k+m)} \right]
\nonumber \\
&=& 0
\end{eqnarray}
This is because we use the plane wave propagator in this model.  
To obtain a finite contribution to the T-odd fragmentation function, we 
must calculate the interference of several diagrams shown in Fig.2 which 
produces the non-trivial phase difference.  
For example, self-energy diagram produces an imaginary part in the present 
kinematics, and thus contributes to the T-odd fragmentation function.  
We find such a contribution from the imaginary part is about $10 \%$ of 
the real part; $ H_1^\bot / D \sim 0.1$.  
The lower diagram of Fig.2 represents the interference of the two-pion 
fragmentation process considered in ref.~\cite{Collins2,JJT}.  
All these diagrams give non-vanishing contributions to the T-odd quark 
fragmentation process.  
Here, instead of doing explicit calculations\cite{Suzuki}, 
we rewrite the quark propagator\cite{Collins} based on the above argument; 
\begin{eqnarray}
S(p) =\frac{A \not p + B}{p^2 - m^2}
\end{eqnarray}
with $A,B$ are complex numbers.  In this case,  we obtain 
\begin{eqnarray}
H _1^\bot (z,k_\bot,s_\bot) =
{{g^2} \over {8\pi ^3}}{z \over {1-z}} 
{ 2 m  \over {\left(
 {k^2-m^2} \right)^2}} \mbox{Im}[A^* B]
\left( {k_\bot \times s_\bot } \right)_z
\label{fragans}
\end{eqnarray}
This clearly shows non-trivial dependence on the transverse spin and 
momentum of the quark.  Calculations presented here are of course 
model-dependent, but the $\left( {k_\bot \times s_\bot } \right)_z$ dependence 
is expected from the T-odd nature of this process.

\section{Single spin asymmetry in inclusive pion production}

In the inclusive pion production in the transversely polarized 
collision $\vec p + p
\to \pi^a + X$, significant analyzing powers are observed at high $x_F$ 
and rather low $p_T \simeq 1 \sim 2$GeV.  
They are also measured for the kaon and eta productions.  
The asymmetry 
grows as $p_T$ increases up to about 2GeV.  Experimental data at higher 
$p_T$ will be available at RHIC or HERA in near future.

Assuming the factorization one can express the inclusive meson 
production cross section as
\begin{eqnarray}
E_h{{d\sigma } \over {d^3p_h}}&=&\sum\limits_{ab} {\int {dx_adx_b{{dz} 
\over z}\;}}q_{a / p}(x_a)\,\rho _{\alpha \alpha' }\;q_{b / p}(x_a) \nonumber\\
&&
\times H_{\alpha \alpha' \beta \beta' }(a+b \to c+d)\;\;D_{\pi / c\;\beta 
\beta '}(z,k_\bot)
\end{eqnarray}
where $q_{a / p}(x_a)$ and $D_{\pi / c\;\beta \beta '}(z,k_\bot)$ are quark 
distribution and fragmentation functions, and $\rho _{\alpha \alpha' }$ the 
helicity density matrix.  
Hard scattering part $H_{\alpha \alpha ' \beta \beta ' }(a+b\to c+d)$ are 
calculated within perturbative QCD.  Using the standard helicity basis, spin 
transfer coefficient is given by\cite{Collins}
\begin{eqnarray}
\frac{M_{++}M_{+-}^* + M_{-+}M_{--}^*}{|M_{++}|^2 + |M_{+-}|^2} 
= \frac{4 (1+\cos \theta)}{4+(1+\cos \theta )^2}
\end{eqnarray}
where $\theta$ is the scattering angle of partons. 
It is easily seen that the spin transfer coefficient 
is large in the forward direction.  
The single spin asymmetry is defined by
$A_N =(d \sigma (\uparrow) - d \sigma (\downarrow)) / 
(d \sigma (\uparrow) + d \sigma (\downarrow))$ .

We are in the position to fix the non-perturbative parts, distribution 
and fragmentation functions.  
For the transversity distribution of the nucleon, 
we use the results of ref.~\cite{KS} by the quark model as inputs.  
We then try to determine the T-odd fragmentation function.  
Here, we simply assume the following form for the T-odd fragmentation 
function;
\begin{eqnarray}
H^\bot_1 (z,k_\bot,s_\bot)=
C {z \over {1-z}} 
{ k^2  \over {\left(
 {k^2-m^2} \right)^2}}
\left( {k_\bot \times s_\bot } \right)_z  \; . 
\end{eqnarray}
which is the simple parametrization motivated by eq.~(\ref{fragans}) in 
section 2.  The constant parameter $C$ is fixed to reproduce the 
$\pi^+$ asymmetry.

\begin{figure}
\begin{minipage}[t]{60mm}
            \epsfxsize = 6.2cm   
            \centerline{\epsfbox{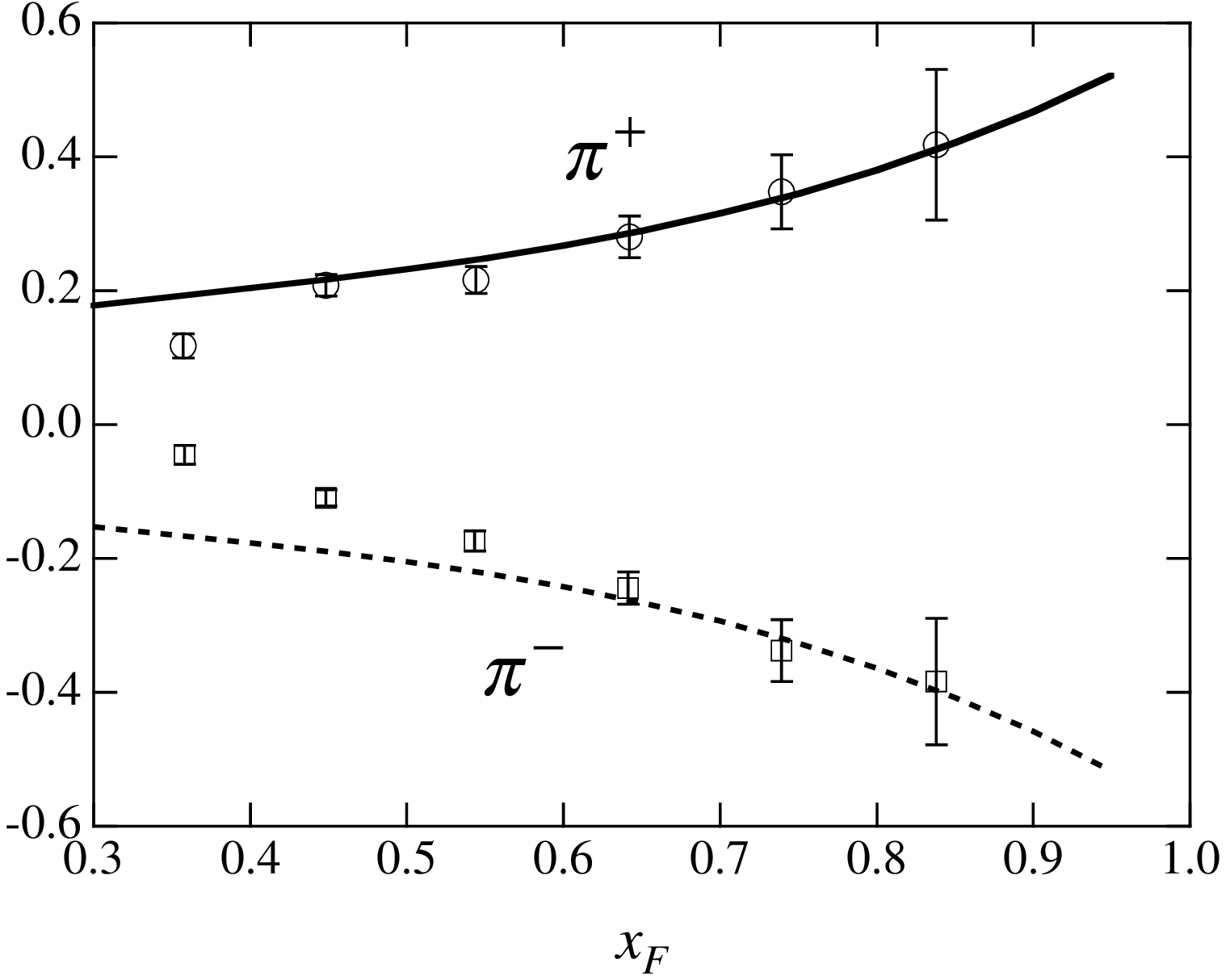}}
          \caption{$A_N$ for $\pi^+$ and $\pi^-$}
          \label{fig3}
\end{minipage}
\hspace{\fill}
\begin{minipage}[t]{60mm}
            \epsfxsize = 6.2cm   
            \centerline{\epsfbox{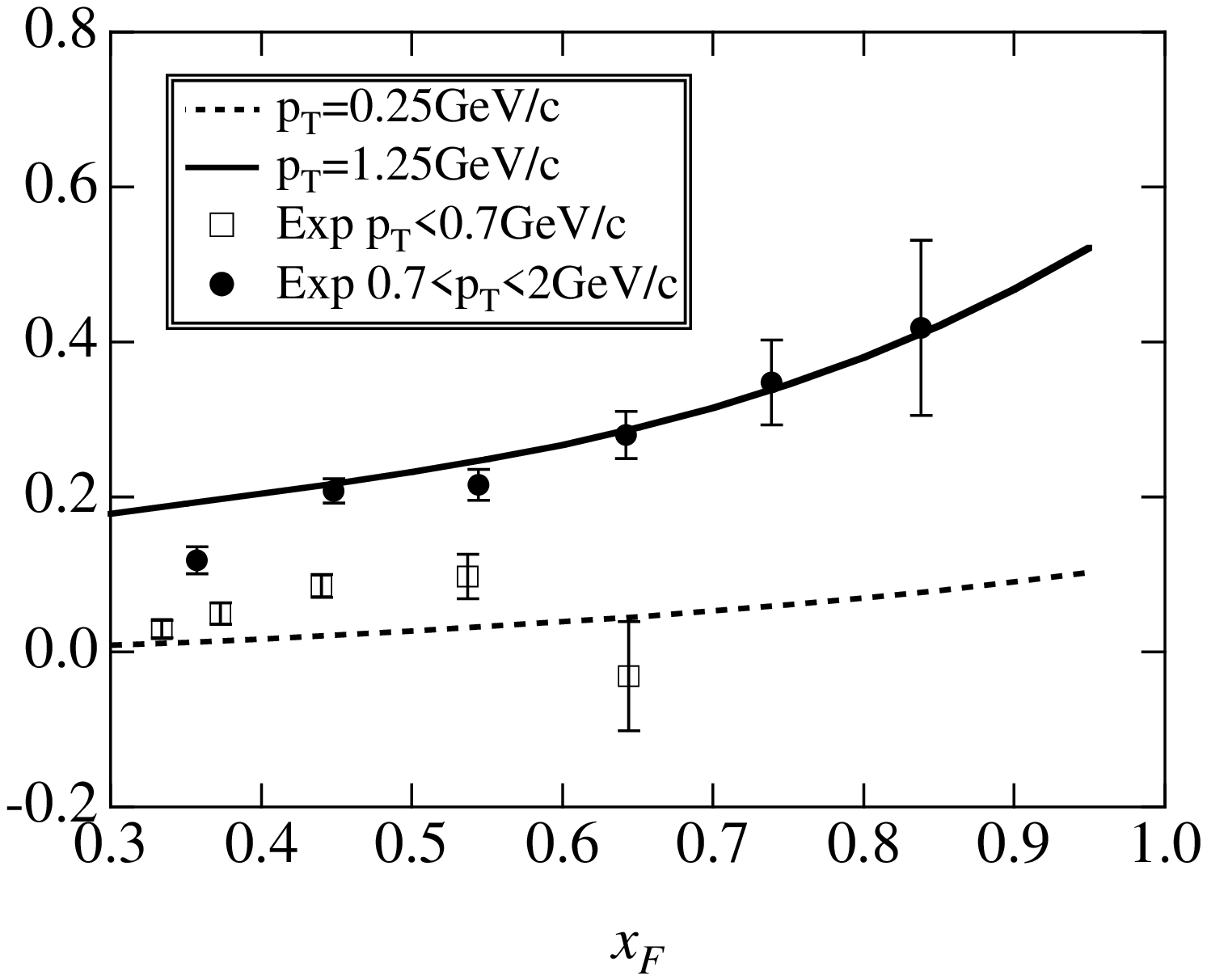}}
          \caption{$p_T$ dependence of $A_N$ for $\pi^+$}
          \label{fig4}
\end{minipage}
\end{figure}

We show in Fig.3 results of the single spin asymmetries for 
$\pi^+$ and $\pi^-$ with the experiments.  Here, Feynman-$x$ is 
roughly approximated by $x_F \sim x_a  z$.  
It is worth noting that high-$x_F$ behavior of the analyzing power is 
sensitive to a ratio of transversity to spin-averaged distributions 
$h_1(x) / f_1(x)$ at the large $x$\cite{SNTK}.  Experimental data 
$A_N (\pi^+) / A_N (\pi^-) \sim -1$ at $x_F \sim 1$ suggests 
$h_1^u(x) / f_1^u (x) \sim - h_1^d(x) / f_1^d (x) \sim 1$ in our 
approach.  
Results for $\pi^0$ as well as $K, \eta$ mesons are also consistent 
with experiments.  
In Fig.4 we show pion $p_T$ dependence of the single spin asymmetry.

\section{Semi-inclusive pion production in DIS}

We apply the T-odd fragmentation function to the deep inelastic scattering 
off a transversely polarized nucleon with a pion in the final state, 
$\ell + \vec p \to \ell ' + \pi^a + X$.  
General analysis of the semi-inclusive process with inclusion of T-odd 
distribution and fragmentation functions is done in 
ref.~\cite{Mulders} at tree level.  
With the unpolarized lepton and the transversely polarized nucleon 
target, one can access the transversity distribution of the nucleon and 
the T-odd quark fragmentation.

Angle $\sin \phi$ 
weighted single spin asymmetry of the pion production cross section 
in ref.~\cite{Mulders} is given by
\begin{eqnarray}
A_{OT}^a&\equiv& \frac{\int d^2 P_\bot \;  \frac{P_{\bot }}
{z MP_{\pi}} \; \sin \phi \; (d \sigma \uparrow - d \sigma \downarrow)}
{\int  d^2 P_\bot \;  (d \sigma \uparrow + d \sigma \downarrow)} 
\nonumber \\
&=& -|S_T| \frac{2 (1-y)} { 1+ (1-y)^2} \frac{h_1^a(x) \; H_1^\bot (z)}
{f_1^a(x) \; D^a (z)}
\label{aot}
\end{eqnarray}
Because we have already determined the T-odd fragmentation function in 
the previous section, 
we can evaluate this asymmetry shown in Fig.5.    
Asymmetry is order of about $10\%$ which is much smaller than the $pp$ 
collision case.  
This is simply due to the difference of kinematical condition.  
In the $pp$ case, the asymmetry at $x_F \sim 1$ is given by the quark 
distributions at $x \sim 1$, in which $h_1(x) / f_1(x) \sim 1$.  
On the other hand, in the semi-inclusive DIS, we observe rather middle or 
small-$x$ behavior of the distribution functions, where 
$h_1(x) / f_1(x)$ is much smaller than unity.

Recently, measurements of the single spin asymmetries were done by 
\cite{HERMES} and \cite{SMC}.  The experimental data actually show 
desired Collins-angle $\sin \phi$ dependence\cite{Boer}.  
Further experiments will clarify the existence of the 
T-odd quark fragmentation process.

Here, we assume that the pion production is dominated by the 
`favored' process like $u \to \pi^+$, and neglect the `unfavored' 
contribution such as $d \to \pi^+$.  Recently, Sch{\"a}fer and Teryaev 
have derived a sum rule for the T-odd fragmentation function\cite{ScTe}, 
and concluded 
that the unfavored process is severely suppressed, which supports our 
assumption adopted here.

\begin{wrapfigure}{r}{6.8cm}   
            \epsfxsize = 6.5cm   
            \centerline{\epsfbox{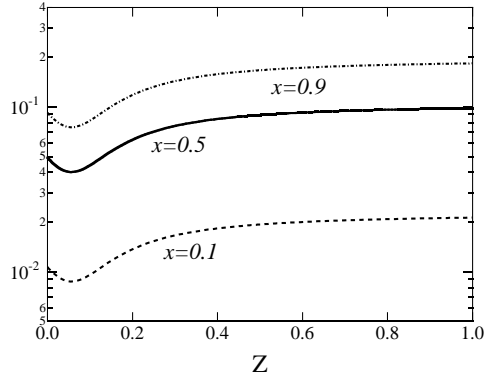}}
          \caption{$A_{OT}$ in semi-inclusive DIS}
          \label{fig:semi}
\end{wrapfigure}

\section{Meson cloud effects on the transversity $h_1 (x)$}

As seen in eq.~(\ref{factor}), the existence of the T-odd 
fragmentation function 
makes it possible to measure the chiral-odd transversity distribution 
function 
$h_1 (x)$ directly.  
In the non-relativistic limit, the transversity distribution coincides 
with 
the helicity one $g_1 (x)$, however, $h_1 (x)$ should differ 
from $g_1 (x)$ in the real world.  
$h_1(x)$ can be understood as the difference of the  
numbers of quarks with eigenvalues $+1$ and $-1$ of the transverse 
Pauli-Lubanski operator $S_\bot \gamma_5$ in the transversely polarized 
nucleon\cite{JJ}.   
Unlike the helicity distribution $g_1(x)$, the gluon operators 
do not mix quark ones.

Simple estimates of the quark model, e.g.~MIT bag model,  yield almost 
similar behavior of the
transversity and helicity distribution functions, although there is a small  
difference\cite{JJ}.  Major source of this difference comes from the 
relativistic effect
on the quark wave function in the nucleon.  
However, interesting observation made in ref.~\cite{KW} 
is that the meson cloud model gives very 
different contributions to $h_1(x)$ and $g_1(x)$.   
One can write the Fock-state of the physical nucleon as
\begin{eqnarray}
\left| p \right\rangle  &=&\sqrt{Z}\left| p_0 \right\rangle + 
a_\pi \left| {n \pi ^+} 
\right\rangle +{{a_\pi } \over 2}\left| {p \pi ^0} \right\rangle + \cdots
\end{eqnarray}
where the renormalization constant $Z$ is found to be  about 0.6 with the 
standard meson-baryon interaction\cite{ST}.  
The quark distribution function of the physical nucleon 
is given by the convolution integral; 
\begin{eqnarray}
q_j (x) =
\int_x^1 {{dy} \over y} \,P_{j \, \pi / i} (y) \, q_i \left( {{x \over y}} 
\right) + \cdots \;\; . 
\label{meson}
\end{eqnarray}
where $P_{j \, \pi / i} (y) $ represents the splitting function 
which gives the probability of the meson-baryon fluctuation 
with the light-cone momentum fraction $y$.  
We calculate the 
splitting functions in the Infinite Momentum Frame with the time-ordered 
perturbation theory.

Let us consider the dominant pion contributions.  
It is well known that the pion cloud provides the $\bar d$ quark excess in the 
nucleon sea, which could account for the violation of Gottfried sum rule.   
For the spin properties of the nucleon, the pion fluctuation also gives 
substantial depolarization effects by emitting the pion into the relative 
$P$-wave state.  
Define the nucleon-pion splitting functions which contribute to 
the spin-averaged, 
helicity, and transversity distributions as $P(y)_{N \pi}$, $\Delta 
P_{N \pi}(y)$ and 
$\delta P_{N \pi} (y)$, respectively.  With the axial-vector pion-nucleon 
interaction, we find a simple exact relation,
\begin{eqnarray}
P(y)_{N \pi} + \Delta P_{N \pi}(y) = 2\delta P_{N \pi} (y) \, . 
\end{eqnarray}
Since $P \ge | \Delta P| ,  | \delta P|$, the pion cloud contributions 
calculated by eq.~(\ref{meson}) are quite different between the 
transversity and helicity distributions.  
With the standard parameters of the meson-baryon interaction\cite{ST}, 
$\Delta P$ is very small and thus  $\delta P$ is about 
a half of $P$ which causes the large effects.

Such a difference is clearly seen in Fig.6.  
Dashed curves show distribution of the bare nucleon (as inputs) and 
the solid ones denote the results with the meson cloud contributions.  
The $d$-quark transversity distribution is suppressed very much compared 
with the helicity one\footnote{For $u$-quark, corrections from 
the meson clouds are small, and essentially negligible.}.  
The 1st moment of the transversity, tensor charge, of $d$-quark is found 
to be reduced by about $40\%$ by the pion depolarization effect, while 
the corresponding axial charge is almost unchanged.  
In fact, such a tendency is consistent with the 
lattice QCD calculation of the tensor charge\cite{Aoki}.

        \begin{figure}
            \epsfxsize = 9cm   
            \centerline{\epsfbox{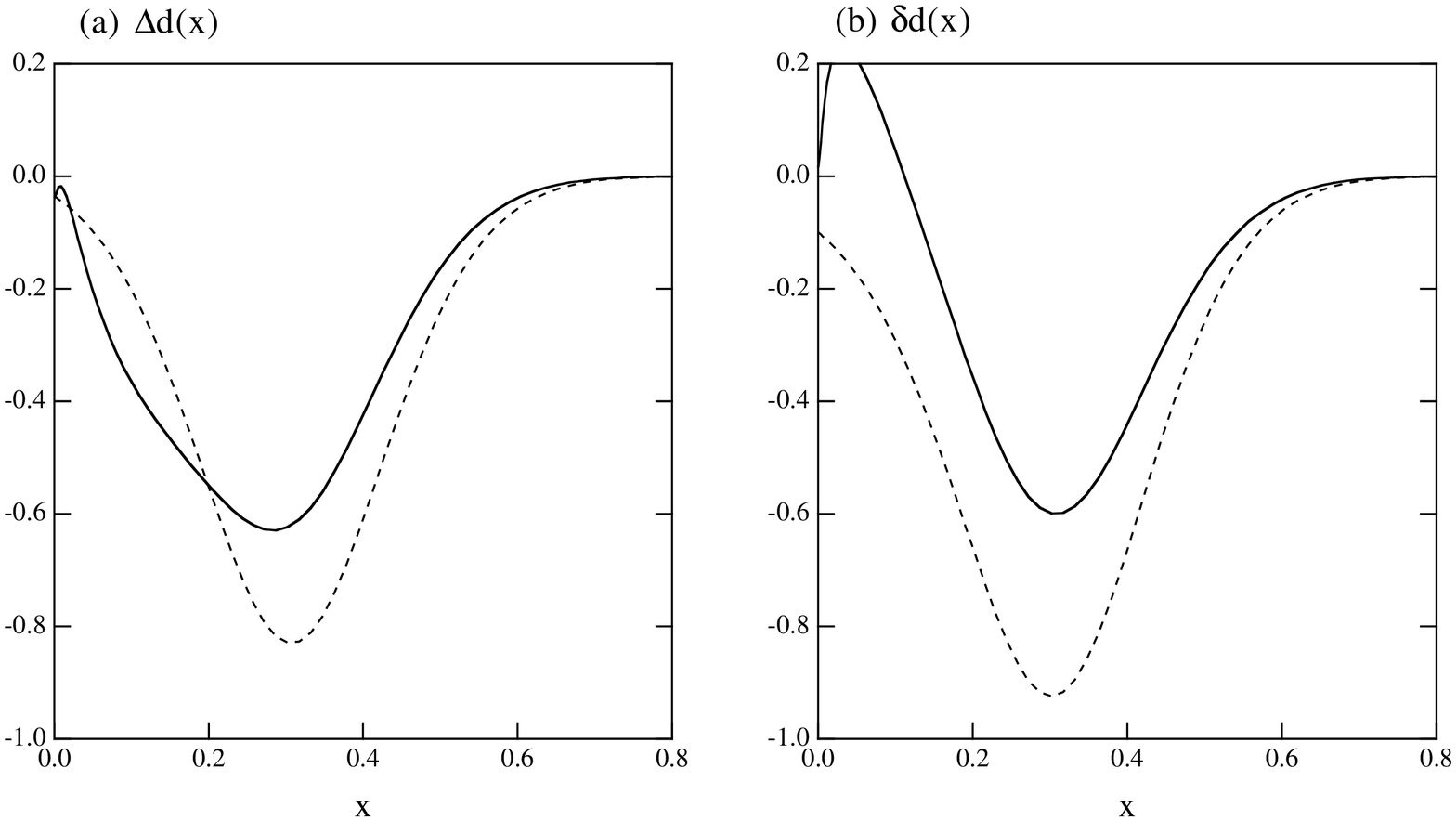}}
        \caption{Modifications of $g_1(x)$ and  $h_1(x)$ of $d$-quark 
in the proton by the pion cloud}
	\end{figure}

In order to clarify the effects of the pion clouds, we take a ratio of 
the $d$-quark distribution to the $u$-quark one.  
In Fig.7 we show our result $h_1^u(x) / h_1^d(x)$ (solid)  
with  $g_1^u(x) / g_1^d(x)$.  
Thick dashed curve denotes our 
calculation for $g_1^d(x) / g_1^u(x)$ with the pion cloud, and thin 
dashed 
one is calculated by the parametrization of ref.~\cite{Brodsky}.  
The ratio of the transversity distributions is very small 
at small-$x$, $x < 0.3$, due to the suppression of the $d$-quark 
transversity distribution.

This ratio can be extracted from the experiments directly.  
Measuring the transverse single spin asymmetries of eq.~(\ref{aot}) for 
charged pions, $\pi^+$ and $\pi^-$, and taking a ratio, one can obtain 
\begin{eqnarray}
\frac{A_{OT}^+}{A_{OT}^-} = \frac{h_1^u(x)}{h_1^d(x)}
\frac{f_1^d(x)}{f_1^u(x)} 
\end{eqnarray}
where the unknown T-odd fragmentation functions are canceled out.  
Hence, one can extract the ratio of transversity $h_1^u(x) / h_1^d(x)$
without ambiguities of the T-odd fragmentation functions.  
Our calculation indicates that the $A_{OT}$ of $\pi^-$ is much 
suppressed compared with $A_{OT}$ of $\pi^+$ and the naive estimate.  
Measurement of the transverse single spin asymmetries for charged pions 
could provide 
severe constraints on the validity of the meson cloud model.

\begin{wrapfigure}{r}{6.8cm}   
            \epsfxsize = 6cm   
            \centerline{\epsfbox{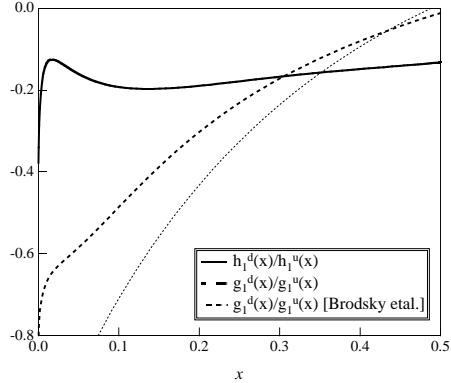}}
        \caption{Modifications of $g_1(x)$ and  $h_1(x)$ of $d$-quark}
          \label{fig:h2}
\end{wrapfigure}

\section{Conclusions}

In summary, we have studied the transverse spin phenomena at high 
energies, namely, T-odd fragmentation function and  transversity 
distribution $h_1(x)$.  
The T-odd quark fragmentation function depends on the transverse spin of 
the quark and thus provides a opportunity to observe $h_1(x)$.  
We have estimated the T-odd quark fragmentation function from 
the analyzing powers in $\vec p + p \to \pi^a +X$ experiments, 
assuming the factorization.  
We then calculate the single spin asymmetry of the semi-inclusive deep 
inelastic scattering off the transversely polarized nucleon 
$\ell + \vec p \to \ell' + \pi^a + X$, which will be measured 
in future experiments.  
We have also discussed the pion cloud effects on the transversity 
distribution $h_1(x)$.   The pion cloud model predicts the strong 
suppression of the $d$-quark transversity distribution of the proton.

\end{document}